\documentclass[sn-mathphys,Numbered]{sn-jnl}

\usepackage{graphicx}%
\usepackage{multirow}%
\usepackage{amsmath,amssymb,amsfonts}%
\usepackage{amsthm}%
\usepackage{mathrsfs}%
\usepackage[title]{appendix}%
\usepackage{xcolor}%
\usepackage{textcomp}%
\usepackage{manyfoot}%
\usepackage{booktabs}%
\usepackage{algorithm}%
\usepackage{algorithmicx}%
\usepackage{algpseudocode}%
\usepackage{listings}%
\usepackage{cleveref}%
\usepackage{tabularx}%
\usepackage{graphbox} 
\usepackage{overpic}
\usepackage{comment}

\begin{document}

\title[Near-Real-Time Mueller Polarimetric Image Processing for Neurosurgical Intervention]{Near-Real-Time Mueller Polarimetric Image Processing for Neurosurgical Intervention}


\author*[1]{\fnm{Stefano} \sur{Moriconi}}\email{stefano.moriconi@insel.ch}

\author[2]{\fnm{Omar} \sur{Rodr{\'i}guez-N{\'u}{\~n}ez}}

\author[3]{\fnm{Romane} \sur{Gros}}

\author[2]{\fnm{Leonard A.} \sur{Felger}}

\author[3]{\fnm{Theoni} \sur{Maragkou}}

\author[4]{\fnm{Ekkehard} \sur{Hewer}}

\author[5]{\fnm{Angelo} \sur{Pierangelo}}

\author[5]{\fnm{Tatiana} \sur{Novikova}}

\author[2]{\fnm{Philippe} \sur{Schucht}}

\author[1]{\fnm{Richard} \sur{McKinley}}

\affil*[1]{\orgdiv{Support Center for Advanced Neuroimaging (SCAN), University Institute of Diagnostic and Interventional Neuroradiology}, \orgname{Inselspital, Bern University Hospital, University of Bern}, \orgaddress{\city{Bern}, \postcode{3010}, \country{Switzerland}}}

\affil[2]{\orgdiv{Department of Neurosurgery}, \orgname{Inselspital, Bern University Hospital, University of Bern}, \orgaddress{\city{Bern}, \postcode{3010}, \country{Switzerland}}}

\affil[3]{\orgdiv{Institute of Tissue Medicine and Pathology}, \orgname{University of Bern}, \orgaddress{\city{Bern}, \postcode{3008}, \country{Switzerland}}}

\affil[4]{\orgdiv{Institute of Pathology}, \orgname{Lausanne University Hospital}, \orgaddress{\city{Lausanne}, \postcode{1011}, \country{Switzerland}}}

\affil[5]{\orgdiv{LPICM}, \orgname{CNRS, Ecole Polytechnique, IP Paris}, \orgaddress{\city{Palaiseau}, \postcode{91120}, \country{France}}}

\abstract{\textbf{Purpose:} Wide-field imaging Mueller polarimetry is a revolutionary, label-free, and non-invasive modality for computer-aided intervention: in neurosurgery it aims to provide visual feedback of white matter fibre bundle orientation from derived parameters.
Conventionally, robust polarimetric parameters are estimated after averaging multiple measurements of intensity for each pair of probing and detected polarised light.
Long multi-shot averaging, however, is not compatible with real-time in-vivo imaging, and the current performance of polarimetric data processing hinders the translation to clinical practice.
\textbf{Methods:} A learning-based denoising framework is tailored for fast, single-shot, noisy acquisitions of polarimetric intensities.
Also, performance-optimised image processing tools are devised for the derivation of clinically relevant parameters.
The combination recovers accurate polarimetric parameters from fast acquisitions with near-real-time performance, under the assumption of pseudo-Gaussian polarimetric acquisition noise.
\textbf{Results:} The denoising framework is trained, validated, and tested on experimental data comprising tumour-free and diseased human brain samples in different conditions.
Accuracy and image quality indices showed significant (p $<$ 0.05) improvements on testing data for a fast single-pass denoising versus the state-of-the-art and high polarimetric image quality standards.
The computational time is reported for the end-to-end processing.
\textbf{Conclusion:} The end-to-end image processing achieved real-time performance for a localised field of view ($\approx$ 6.5 mm$^2$).
The denoised polarimetric intensities produced visibly clear directional patterns of neuronal fibre tracts in line with reference polarimetric image quality standards; directional disruption was kept in case of neoplastic lesions.
The presented advances pave the way towards feasible oncological neurosurgical translations of novel, label free, interventional feedback.}


\maketitle

\section{Introduction}
\label{Intro}
Many characteristics of biological tissues are reflected in their optical properties. 
Differences in birefringence, i.e. the speed of light through that medium depending on polarisation, may change in healthy and diseased tissues.
Mueller polarimetric imaging (MPI) non-invasively measures these optical properties, providing micro-structural features of a sample without contrast media \cite{Qi2017Mueller,Li2022Polarimetric, Ramella-Roman2020Review}.
Intensity images of the superficial layer are acquired by shining light at different polarisation states.
The back-scattered light is then captured by an optical sensor operating in reflection \cite{Baba2002Development} configuration.
As per the polarimetric Stokes-Mueller formalism \cite{Azzam2016Stokes-vector}, the tissue-specific Mueller coefficients are derived from the acquired intensities by solving a linear system.
Polarimetric parameters, including retardance (which characterises the anisotropy of the refractive index of a sample), diattenuation and depolarisation, are determined via different decompositions of the Mueller matrix \cite{Li2021Polaromics,Lu1996Interpretation,SanJose2023Extended}.
In diagnostic clinical applications, MPI identified disease progression by revealing morphological tissue changes \textit{ex-vivo} \cite{Pierangelo2011ExVivo,Rehbinder2016Mueller}.
In \cite{Axer2001Quantitative,Axer2011HighResolution}, polarised light first estimated the neuronal fibre bundle orientation in histological sections of formalin-fixed human brain, towards a tractographic reconstruction of the white matter as in diffusion weighted MRI.
In \cite{Schucht2020Visualisation}, a wide-field MPI system showed white matter fiber tracts on fresh and formalin-fixed samples of different specimens, paving the way towards label-free neurooncological visualisations.
In the aforementioned studies, real-time performance was not initially sought, since high image quality was prioritised to accurately characterise the samples' properties.
The accurate estimation of the azimuth of the optical axis, indicative of the orientation of fibre bundles within the imaging plane, is key for neurosurgery \cite{RodriguezNunez2021Retardance}.
Under the assumption that lesions alter the organised arrangement of neuronal fibres, directional cues in axonal pathways may guide the resections of neoplastic tissue in white matter.
Neurosurgeons would be informed on tumour boundaries and surrounding healthy tissues, irrespective of tissue ablation and displacement, beyond available navigation systems based on preoperative image planning.
High accuracy is also key to discriminate among different tissues, such as neoplastic types and grades showing different degrees of infiltration \cite{Qi2017Mueller}, and to perform tissue classification tasks leveraging artificial intelligence (AI) techniques \cite{McKinley2022Machine}.
With this view, MPI acquisition noise and latency represent two main bottlenecks for the accurate image processing, quantitative analysis, and ultimate neurosurgical feedback.
The enhancement of polarimetric image contrast traditionally requires long-time, multi-shots, averaging techniques to reduce the acquisition noise of the optical system and sensor camera \cite{Novikova2023Mueller}.
Furthermore, computational noise propagates from unfiltered acquisitions throughout the derivation of polarimetric parameters in the cascade of numeric Mueller decompositions \cite{SanJose2023Extended,Ossikovski2008Product}.
An optimal MPI enhancement embedded in a performance-optimised image processing pipeline is a key-enabling technology for revolutionising computer-aided neurosurgery, and currently stand as an open translational challenge.

\subsection{Contribution and Outline}
\label{Intro:AimContributions}
Aiming to tackle the aforementioned challenges, in this feasibility study we introduce an AI-based framework integrated with performance-optimised polarimetric image processing tools, to simultaneously denoise low quality single-shot polarimetric acquisitions and to boost the performance and the estimation of relevant parameters in an end-to-end pipeline.
A denoising diffusion network tailored for polarimetric intensity data is introduced in \cref{Methods:PolarimDenoisingDiffusionNetwork}, and performance-optimised polarimetric image processing tools are described in \cref{Methods:PolarimComputationalToolkit}. 
Experiments validate the proposed framework for real polarimetric images of human brain tissues in different conditions in \cref{ExperimResults}.
Observations on neurosurgical MPI are discussed in \cref{DiscussionConclusions}.

\section{Polarimetric Denoising Diffusion Network}
\label{Methods:PolarimDenoisingDiffusionNetwork}

\begin{figure}[t]
    \centering
    \includegraphics[width=.9\textwidth]{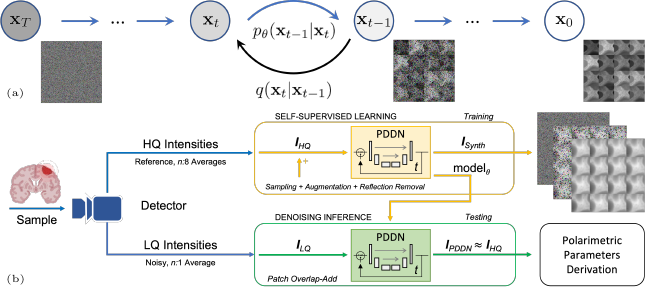}
    \caption{Denoising diffusion model. (a) Forward and reverse diffusion as in \cite{Ho2020Denoising}. Degraded states $\mathbf{x}_t$ of the forward diffusion (black arrow) in the modelled Markov chain of $T$ time-points. Inferential sampling of the reverse diffusion (blue arrow) over the parametrised distribution $p_{\theta}(\mathbf{x}_{t-1}|\mathbf{x}_t)$. (b) Schematic diagram of the AI-based denoising polarimetric framework. PDDN builds on a time-point recursive U-Net for the reverse diffusion, as in \cite{DDN2022GitHub}.}
    \label{fig:MarkovChainDiffusion}
\end{figure}
The denoising of a short-time, low-quality, single-shot polarimetric acquisition is performed with a deep-learning implementation of a diffusion probabilistic model \cite{SohlDickstein2015Deep,Ho2020Denoising}.
Diffusion probabilistic models can generate images as the composition of many small \textit{denoising} steps.
Our polarimetric denoising diffusion network (PDDN) is a tailored adaptation of this framework, leveraged to literally reduce the acquisition noise from polarisation state intensities with high-performance.
In a probabilistic diffusion process, an input image $\mathbf{x}_{0}$ is gradually corrupted with additional Gaussian noise over a series of $T$ time-points.
Such degradation process, i.e. forward diffusion, determines for each $T-1$ adjacent time-step a pair of images: a degraded instance and less-degraded one.
In the reverse diffusion, the ill-posed recovery of clean data from noisy instances is achieved for each time-step by reversing the degradation process with a neural network for conditional inference.
Here, the noise level of short-time, single-shot polarimetric scans is assumed comparable to the degradation at specific time-points in a forward diffusion, so that the image restoration is parametrised with a progressively small recovery, as in the reverse diffusion.
As in \cite{Ho2020Denoising}, the diffusion is modelled with a Markov chain of $T$ time-points, where the image state at each time-point $t$ only depends on the image state at $t-1$.
For a clean input image $\mathbf{x}_0$ at $t=0$, the full degrading trajectory to $\mathbf{x}_T$ in \cref{fig:MarkovChainDiffusion} is formulated as the sequential product of the posterior probability $q(\mathbf{x}_t|\mathbf{x}_{t-1}) = \mathcal{N}(\mathbf{x}_t;\boldsymbol{\mu}_t=\sqrt{1-\beta_t}\mathbf{x}_{t-1},\boldsymbol{\Sigma}_t=\beta_t\mathbf{J})$ for each degrading time-step, with $\boldsymbol{\mu}_t$ and $\boldsymbol{\Sigma}_t$ the tensorial mean and variance of the Gaussian noise with scalar time-dependent variance $\beta_t$, and $\mathbf{J}$ the identity matrix.
To efficiently sample any degradation state $\mathbf{x}_t$ from $\mathbf{x}_0$, a closed form is obtained by re-parametrising the scalar variance $\beta_t$ leveraging a canonical Gaussian noise $\boldsymbol{\epsilon} \sim \mathcal{N}(\boldsymbol{0},\mathbf{J})$, as in \cite{Ho2020Denoising}.
In the reverse diffusion, the estimation of the distribution $q(\mathbf{x}_{t-1}|\mathbf{x}_t)$ as in \cref{fig:MarkovChainDiffusion} is approximated assuming an underlying Markov chain of $T\rightarrow\infty$ time-points, and additional Gaussian noise with a small scalar variance $\beta_t$ at each time-point. 
The parametric formulation of the reverse trajectory therefore approximates the distribution as $p_{\theta}(\mathbf{x}_{t-1}|\mathbf{x}_t) = \mathcal{N}(\mathbf{x}_t;\boldsymbol{\mu}_{\theta}(\mathbf{x}_t,t),\boldsymbol{\Sigma}_{\theta}(\mathbf{x}_t,t))$, where a learning paradigm regresses the tensorial mean $\boldsymbol{\mu}_{\theta}(\mathbf{x}_t,t)$ and variance $\boldsymbol{\Sigma}_{\theta}(\mathbf{x}_t,t)$ from pairs of image states, for each time-point, optimising a loss derived from the evidence lower bound, in the form of a negative log-likelihood.
As in \cite{Ho2020Denoising}, a simplified formulation of the loss accounts for the re-parametrisation of the additional noise variance, assumed identical in each tensorial dimension, with further conditioning on the input image.
The joint distribution of the reverse diffusion is translated into an encoder-decoder coupling as in a U-Net \cite{Ronneberger2015U-Net}, where denoising kernels are learned in a self-supervised fashion.
As in \cref{fig:MarkovChainDiffusion}, the PDDN is trained on unpaired high-quality intensities obtained from long-time averaged acquisitions.
At convergence, the model denoises short-time, low-quality, single-shot polarimetric images for few terminal steps, at inference.
\paragraph{Network Implementation Details.}
\label{Methods:NetworkImplementDetails}
The PDDN implements a time-point recursive U-Net\cite{Ronneberger2015U-Net}, as in \cite{DDN2022GitHub}.
Four deep layers of wide ResNet blocks, group normalization and self-attention blocks are employed, with pooling and upsampling scheme of $(1,2,4,8)$, each with $3 \times 3$ convolutional kernels, unitary stride and sigmoid linear units (SiLU) activation function, alternated by skip connections between the encoder-decoder branches. 
A total of $T=1000$ time-points were considered.
The training used an L1-loss with an Adam optimiser, learning rate $l_{r} = 1e-4$ and 100k epochs, with batches of 32 sampled and augmented patches of data.
Full model memory footprint: 250MB.
Polarimetric intensities were arranged in tensor patches of size $128\times128\times16$, with the first two dimensions encompassing spatial extent and the third the fixed polarimetric measurement states.
Data augmentation included random rotation, flip and cropping with mirroring padding.
Supra-threshold intensity reflections were masked to avoid spurious artifacts and hallucinations.
Intensities were linearly re-scaled within $[-1,1]$.

\section{Efficient calculation of Polarimetric Biomarkers}
\label{Methods:PolarimComputationalToolkit}
\begin{figure}[t]
    \centering
    \includegraphics[width=\textwidth]{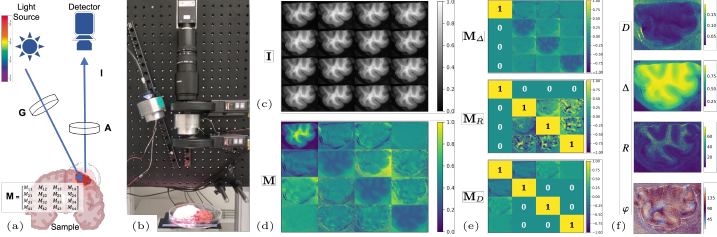}
    \caption{Wide-field MPI system in reflection configuration. (a) Schematic: light source, polarisation state generator $\mathbf{G}$, biological sample, polarisation state analyser $\mathbf{A}$, and detecting camera. (b) Our instrumentation. (c) Acquired polarisation states intensity image $\mathbf{I}$. (d) Derived full Mueller matrix $\mathbf{M}$. (e) Lu-Chipman decomposition: $\mathbf{M}_{\Delta}$, $\mathbf{M}_{R}$, and $\mathbf{M}_{D}$ matrices. (f) Derived scalar parameters: $D$, $\Delta$, $R$, and $\varphi$. Enhanced contrast of $\mathbf{M}$ with a sigmoid mapping.}
    \label{fig:widefieldMPI}
\end{figure}
Divide-and-conquer approaches are employed together with linear algebra vectorisation and parallel computing to boost the performance of Mueller matrix decomposition and the extraction of accurate polarimetric parameters.
In the developed performance-optimised image processing tools, polarimetric derivations, as in \cref{fig:widefieldMPI}, are formulated as a system of linear equations \cite{Azzam2016Stokes-vector}. 
For a MPI system in reflection configuration, the Mueller matrix is derived from a 16-channel tensor of 2D intensities of size $H \times W$ pixels, for all $(4 \times 4)$ polarisation states as
\small
\begin{equation}
    \label{eq:LinSystem}
    \mathbf{M}=\mathbf{A}^{-1}\,\mathbf{I}\,\mathbf{G}^{-1},
\end{equation}
\normalsize
where $\mathbf{M}$ is the unknown full $4 \times 4$ Mueller matrix of each pixel, $\mathbf{G}$ and $\mathbf{A}$ are the pixel-wise $4 \times 4$ matrices of the polarisation state generator and analyser, respectively determined at calibration, and $\mathbf{I}$ is the $4 \times 4$ tensor of real-valued intensities measured by the camera for the considered pixel, where each component accounts for a different combination of the elicited polarisation states.
The linear system in \cref{eq:LinSystem} can be solved in closed form for each Mueller coefficient as sum of scalar products and represents an explicit vectorisation of the solution for tensors of arbitrary dimensions.
Such formulation pixel-wise solves for the Mueller matrix coefficients, where parallel computing enables high-dimensional vectorised data processing with arbitrary hardware capacity.
Following \cite{Lu1996Interpretation} the Mueller matrix is decomposed as the matricial product of three optical components, i.e., the diattenuator, the retarder and the depolariser, as $\mathbf{M}=\mathbf{M}_{\Delta}\,\mathbf{M}_{R}\,\mathbf{M}_{D}$.
Scalar maps of polarimetric parameters are pixel-wise derived from decomposed polarimetric tensors, accounting for total diattenuation ($D$), total depolarisation ($\Delta$), scalar retardance ($R$), and the azimuth of optical axis ($\varphi$) as
\small
\begin{gather}
    D = \sqrt{M_{D_{12}}^{2} + M_{D_{13}}^{2} + M_{D_{14}}^{2}} ~~ \text{and} ~~ \Delta = 1-\frac{1}{3}\vert tr(\mathbf{M}_{\Delta})\vert, \\
    R = \cos^{-1} \left(\sqrt{ (M_{R_{22}}+M_{R_{33}})^2+(M_{R_{32}}-M_{R_{23}})^2}-1\right) ~ \text{and} ~
    \varphi = \frac{1}{2}\tan^{{-}1} \left (\frac{M_{R_{24}}}{M_{R_{43}}}\right).
\end{gather}
\normalsize
All vectorisations are implemented with compiled routines, and all derivations are wrapped in scripting languages for high-level development of AI designs \cite{libmpMuelMat2022GitHub}.
\section{Experiments and Results}
\label{ExperimResults}
\paragraph{Data.}
\label{ExperimResults:Dataset}
Polarimetric data $\mathbf{I}$ were acquired with a wide-field imaging Mueller polarimeter as in \cite{Novikova2023Mueller} \cref{fig:widefieldMPI}, at 550 nm wavelength, with a CCD camera (Stingray F080B, Allied Vision, Germany) $512\times384$ pixels ($20\times24$ mm FoV, resulting in $\approx 50 \mu$m resolution).
Matrices $\mathbf{G}$ and $\mathbf{A}$ in eq. \eqref{eq:LinSystem} were determined at calibration.

\noindent \textit{Training.} $~$ PDDN was trained on 200 high-quality (HQ) images from multi-shots averaged ($n=8$) acquisitions of \textit{fresh} human brain tissues from neurosurgical resections and post-mortem examinations. Portions of grey and white matter were resected from cortical regions involving eloquent areas of the brain. These included tumour-free and neoplastic samples ($\approx 50\%$ ratio) of different types (gliomas, meningiomas, metastases), at varying degrees of severity and infiltration.

\noindent \textit{Validating.} $~$ 50 HQ images of mixed fresh and formalin-fixed brain tissues from healthy human and animal specimens exhibited similar contrast and minor biological heterogeneity (cortical and deep-brain structures) for the model optimisation.

\noindent \textit{Testing.} $~$ A different set of 200 rigidly co-registered paired images of \textit{mixed} fresh and formalin-fixed human brain samples including tumour-free and neoplastic tissues ($\approx 50\%$ ratio) were acquired at low-quality (LQ), i.e. short-time, noisy, single-shot ($n=1$), and at HQ, respectively.
Representative annotated data were acquired also at super high-quality (SHQ): multi-shot ($n=16$) averaged acquisitions, for a case study.
\paragraph{Evaluation.}
\label{ExperimResults:QuantitativeEval}
The accuracy and the performance of the denoising framework were compared to traditional approaches, alternative methods and the state-of-the-art.
The evaluated polarimetric instances comprised denoised $\mathbf{I}$ and derived $\mathbf{M}$, as well as the scalar maps $D$, $\Delta$, $R$, and $\varphi$.
Image quality scores including the root-mean-squared error (RMSE), the normalised peak signal-to-noise ratio (nPSNR) and the structural similarity index (SSIM) were pixel-wise computed for the paired test data, as in \cite{Yang2022DeepLearning}.
Values and deviations of angular data ($R$ and $\varphi \in [0, \pi]$), were computed with circular statistics and reported in degrees.
Scores were evaluated within a region of interest (ROI) separating tissues from background.
Reflection artifacts were excluded from the analysis.
Significant differences were assessed with a pairwise Wilcoxon rank sum test \cite{Gibbons2014NonParametric}.

\subsection{Denoising Polarimetric Intensities of Human Brain Tissues}
\label{ExperimResults:DenoisingPolarimIntensities}

\begin{figure}[!t]
    \centering
    \includegraphics[width=.9\textwidth]{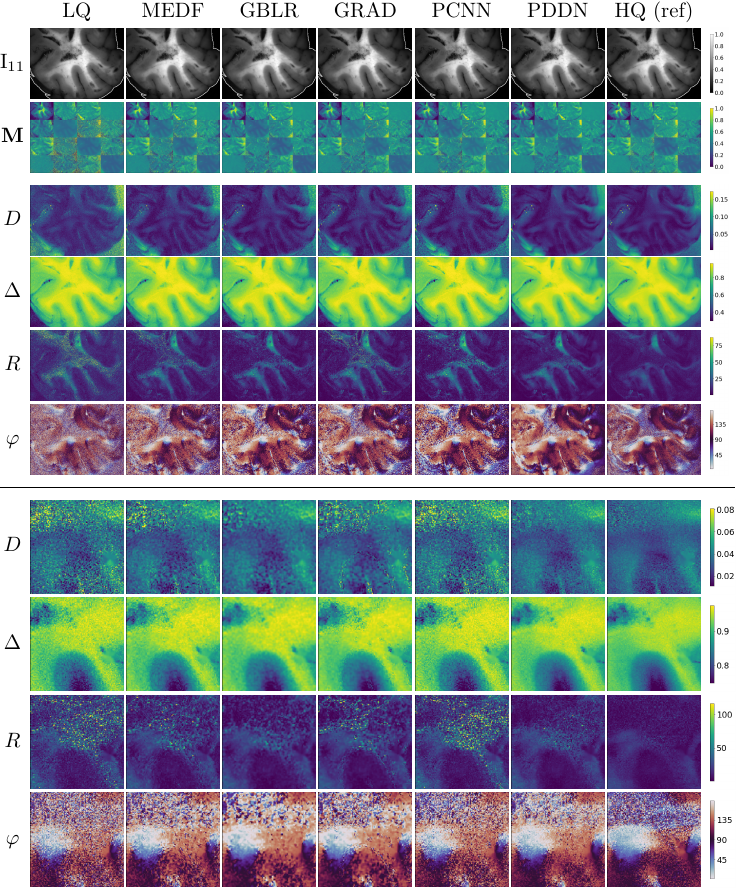}
    \caption{Denoising Polarimetric Intensities of Human Brain Tissues: gallery of polarimetric instances. (top) \textit{Tumour-free} sample of the \textit{testing} set. $\text{I}_{11}$ component shown with the evaluation ROI contour. (bottom) Details of polarimetric parameters in a centre-cropped area. Images in each row have the same range of values as in the colour-bar. $R$ and $\varphi$ reported in degrees.}
    \label{fig:Results_healthy}
\end{figure}

\begin{figure}[!t]
    \centering
    \includegraphics[width=.9\textwidth]{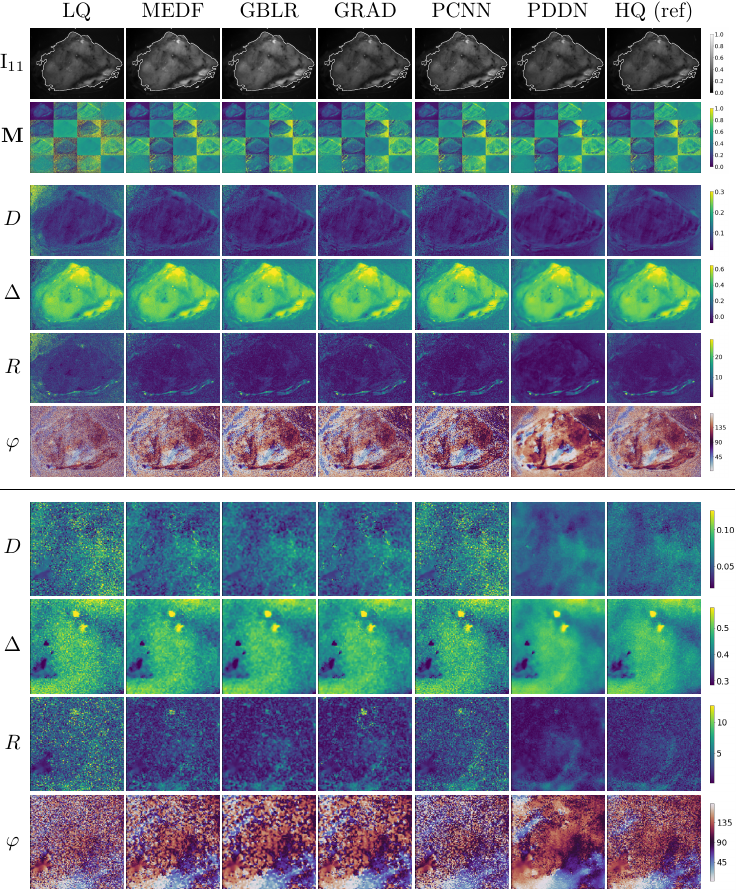}
    \caption{Denoising Polarimetric Intensities of Human Brain Tissues: gallery of polarimetric instances. (top) \textit{Neoplastic lesion} of the \textit{testing} set. $\text{I}_{11}$ component shown with the evaluation ROI contour. (bottom) Details of polarimetric parameters in a centre-cropped area. Images in each row have the same range of values as in the colour-bar. $R$ and $\varphi$ reported in degrees.}
    \label{fig:Results_diseased}
\end{figure}

Low-quality intensity images of polarimetric states of the testing set were denoised with a single-pass ($t=1$) PDDN filtering step and with a set of traditional and AI-based methods.
Traditional averaging schemes yielded HQ images as reference ground-truth.
State-of-the-art polarimetric convolutional denoising networks (PCNN) \cite{Li2020LearningBased,Yang2022DeepLearning} were trained on paired intensity instances of the training set.
Also, deterministic denosing algorithms were considered as baseline: the median filter (MEDF: 3-kernel)  \cite{Huang1979Fast}, the Gaussian blur (GBLR: 5-kernel) \cite{Buades2005Non}, and the gradient anisotropic diffusion (GRAD: 5 steps, 1 conductance) \cite{Krissian2007Oriented}.
Qualitative results are shown in \cref{fig:Results_healthy,fig:Results_diseased} for representative tumour-free and neoplastic cases.
In general, minor and subtle changes are observed in the processed intensities of polarisation states and derived Mueller matrices.
Conversely, the effect of denoising was predominant on derived polarimetric parameters of clinical relevance.
Successful denoising is obtained for PDDN, PCNN, and GBLR with improved rejection of acquisition noise compared to LQ acquisitions.
Limited denoising is found for MEDF and GRAD, where noisy patterns remained visible, and the physical characterisation of the underlying sample remained unclear or partially altered.
Overall, polarimetric parameters showed high sensitivity to acquisition noise, to the computational error propagation, and to the denoising method of choice, where $D$, $R$, and $\varphi$ exhibited major deviations between noisy and processed instances.
Image quality scores in \cref{tab:IntensityData_ImageQualityScores} supported the qualitative analysis, where the proposed PDDN reported best values in all cases for all indices, followed by PCNN and GBLR.
This suggests the early and optimal rejection of subtle acquisition noise with PDDN is effective to reduce further error propagation in the computational cascade of polarimetric parameters.
Major deviations were found against LQ instances, where oriented patterns of white matter fibres in $\varphi$ can only be clearly observed after denoising.
The significant improvements obtained with PDDN in \cref{tab:IntensityData_ImageQualityScores} suggest the learned filtering kernels in the proposed polarimetric denoising diffusion network are suitable for enhancing the image quality with minimal deviations compared to reference HQ data using a fast, single-pass, filtering step.

\begin{table}[b]
    \centering
    \tiny
    \begin{tabular}{@{}c@{\hskip .5em}c@{\hskip .5em}c@{\hskip 1em}c@{\hskip 1em}c@{\hskip 1em}c@{\hskip 1em}c@{\hskip 1em}c@{}}
        & Modality &$\mathbf{I}$&$\mathbf{M}$&$D$&$\Delta$&$R$&$\varphi$\\
        \toprule
        \multirow{6}{*}{\rotatebox[origin=c]{90}{RMSE $\downarrow$}}
		& LQ & 0.9/1.1/1.5 & 0.8/0.9$^{\ast}$/1.1 & 1.4/1.6$^{\ast}$/1.9 & 2.4/2.8$^{\ast}$/3.4 & 5.4/9.1$^{\ast}$/15.5 & 31.0/35.7$^{\ast}$/41.7 \\
		& MEDF & 0.9/1.1/1.5 & 0.7/0.8$^{\ast}$/0.9 & 0.9/1.2$^{\ast}$/1.3 & 1.8/2.1$^{\ast}$/2.7 & 3.8/5.7$^{\ast}$/8.7 & 25.7/31.4$^{\ast}$/37.6 \\
		& GBLR & 0.9/1.2$^{\ast}$/1.6 & 0.6/0.7$^{\ast}$/0.8 & 0.8/0.9$^{\ast}$/1.1 & 1.5/1.9$^{\ast}$/2.4 & 3.1/4.4/6.7 & 22.4/28.5$^{\ast}$/35.3 \\
		& GRAD & 0.9/1.1/1.5 & 0.8/0.9$^{\ast}$/1.0 & 1.2/1.4$^{\ast}$/1.7 & 2.1/2.5$^{\ast}$/3.0 & 4.8/8.3$^{\ast}$/12.3 & 27.3/32.7$^{\ast}$/38.3 \\
		& PCNN & 1.6/1.9$^{\ast}$/2.5 & 0.7/0.8$^{\ast}$/0.9 & 0.8/1.0$^{\ast}$/1.1 & 1.7/2.1$^{\ast}$/2.7 & 2.9/4.5/6.7 & 23.3/28.9$^{\ast}$/35.1 \\
		& \textbf{PDDN} & \textbf{0.7}/\textbf{1.00}/\textbf{1.4} & \textbf{0.5}/\textbf{0.6}/\textbf{0.7} & \textbf{0.7}/\textbf{0.8}/\textbf{0.9} & \textbf{1.4}/\textbf{1.7}/\textbf{2.1} & \textbf{2.5}/\textbf{4.0}/\textbf{6.3} & \textbf{18.9}/\textbf{25.4}/\textbf{32.5} \\
	\midrule
        \multirow{6}{*}{\rotatebox[origin=c]{90}{nPSNR $\uparrow$}}
		& LQ & 28.2/32.1$^{\ast}$/34.1 & 27.5/32.2/35.2 & 6.7/7.6$^{\ast}$/8.6 & 26.7/29.6$^{\ast}$/31.2 & -2.1/0.5$^{\ast}$/4.2 & 9.4/10.9$^{\ast}$/12.4 \\
		& MEDF & 27.7/32.1/34.5 & 26.9/31.6/34.4 & 9.6/10.5$^{\ast}$/11.6 & 28.9/32.1$^{\ast}$/33.7 & 2.3/4.3$^{\ast}$/7.3 & 10.2/12.1$^{\ast}$/13.9 \\
		& GBLR & 27.1/31.0$^{\ast}$/33.6 & 26.3/30.5$^{\ast}$/33.4 & 11.4/12.4$^{\ast}$/13.5 & 30.1/32.9/35.4 & 4.4/6.3/8.8 & 10.6/12.8$^{\ast}$/15.1 \\
		& GRAD & 28.2/32.3/34.3 & 27.5/31.8/34.9 & 7.9/8.9$^{\ast}$/10.0 & 27.9/30.8$^{\ast}$/32.3 & -0.36/2.00$^{\ast}$/5.5 & 9.9/11.7$^{\ast}$/13.5 \\
		& PCNN & 22.9/26.5$^{\ast}$/28.7 & 22.2/26.3$^{\ast}$/28.4 & 10.8/11.8$^{\ast}$/12.9 & 28.7/32.1$^{\ast}$/34.2 & 3.9/6.4$^{\ast}$/8.8 & 10.6/12.8$^{\ast}$/14.8 \\
		& \textbf{PDDN} & \textbf{28.7}/\textbf{33.2}/\textbf{35.7} & \textbf{27.5}/\textbf{32.3}/\textbf{35.3} & \textbf{12.5}/\textbf{13.4}/\textbf{14.4} & \textbf{30.8}/\textbf{33.9}/\textbf{36.1} & \textbf{4.9}/\textbf{7.3}/\textbf{10.0} & \textbf{11.3}/\textbf{13.9}/\textbf{16.5} \\
	\midrule
        \multirow{6}{*}{\rotatebox[origin=c]{90}{SSIM $\uparrow$}}
		& LQ & 99.5/99.8/99.9 & 99.3/99.5$^{\ast}$/99.6 & 49.2/57.5$^{\ast}$/65.1 & 91.8/94.2$^{\ast}$/96.2 & 20.5/34.3$^{\ast}$/61.5 & 17.9/33.3$^{\ast}$/47.2 \\
		& MEDF & 99.6/99.8/99.9 & 99.5/99.7$^{\ast}$/99.8 & 64.8/71.5$^{\ast}$/77.8 & 94.9/96.6$^{\ast}$/97.7 & 37.2/57.2$^{\ast}$/73.1 & 26.9/44.7$^{\ast}$/58.8 \\
		& GBLR & 99.5/99.8$^{\ast}$/99.9 & 99.6/99.8$^{\ast}$/99.8 & 73.3/79.4$^{\ast}$/84.5 & 95.9/97.5$^{\ast}$/98.3 & 48.5/68.0/82.5 & 30.5/50.1/67.7 \\
		& GRAD & 99.5/99.8/99.9 & 99.4/99.5$^{\ast}$/99.7 & 55.4/63.2$^{\ast}$/70.4 & 93.7/95.5$^{\ast}$/97.1 & 23.2/42.0$^{\ast}$/66.7 & 25.1/41.1$^{\ast}$/53.9 \\
		& PCNN & 98.9/99.5$^{\ast}$/99.7 & 99.3/99.6$^{\ast}$/99.7 & 72.3/78.5$^{\ast}$/83.4 & 94.9/96.9$^{\ast}$/97.9 & 48.6/69.5/82.1 & 28.6/45.7$^{\ast}$/61.8 \\
		& \textbf{PDDN} & \textbf{99.6}/\textbf{99.8}/\textbf{99.9} & \textbf{99.7}/\textbf{99.8}/\textbf{99.8} & \textbf{77.4}/\textbf{83.5}/\textbf{87.5} & \textbf{96.7}/\textbf{98.0}/\textbf{98.7} & \textbf{50.5}/\textbf{71.9}/\textbf{85.9} & \textbf{34.7}/\textbf{55.1}/\textbf{73.1} \\
	\botrule
    \end{tabular}
    \normalsize
    \caption{Image Quality Scores: human brain tissues. Denoising $\mathbf{I}$ and evaluating derived parameters. Quartiles: q$_1$/median/q$_3$. Comparison of denoising modalities; multi-shot averaging HQ images as reference. Scores: nPSNR [dB] and SSIM [$\%$]. RMSE scores of $\mathbf{I}$, $\mathbf{M}$, $D$ and $\Delta$ are multiplied by 1e-2. Best values in bold. Significant differences from PDDN: $\ast$ = $p$-value $<0.05$, pairwise Wilcoxon rank sum test.}
    \label{tab:IntensityData_ImageQualityScores}
\end{table}

\subsection{Fibre Orientation and Azimuth Variation}

\begin{figure}
    \centering
    \includegraphics[width=\textwidth]{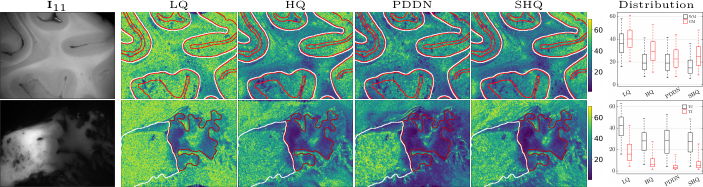}
    \caption{Azimuth $\varphi$ Variations and Distributions: intensity and circular standard deviation \textit{csd($\varphi$)} in degrees. Low values of \textit{csd($\varphi$)} for homogeneous directional patterns, whereas high values for high disruption of fibres orientation or change of directional patterns. (top) \textit{Tumour-free} sample: variability of fibres orientations, \textit{csd} distributions in annotated White (WM) and grey (GM) matter. (bottom) \textit{Neoplastic lesion}: variability of fibres orientations in diseased white matter, \textit{csd} distributions in annotated Tumour Centre (TC) and Infiltration (TI). High rejection of background noise, and visually comparable directional patterns of the fibres after denoising, similarly to high quality image standards. Boxplots: consistent \textit{csd} drop in PDDN as in HQ and SHQ. Better PDDN separation in tumour areas.}
    \label{fig:SanityCheckAzimuthalCircSD}
\end{figure}

Assuming lesions alter the directional arrangement of fibres in white matter, we focus on PDDN and evaluate the effect of denoising on the derived $\varphi$ as a sanity check in tumour-free brain tissue and in a glioma lesion.
The directional variability is evaluated with the azimuth circular standard deviation, i.e. \textit{csd($\varphi$)} in a 5x5 image pixel neighbourhood.
Low \textit{csd($\varphi$)} indicates homogeneous directional patterns, whereas high \textit{csd($\varphi$)} corresponds to degrees of directional disruption.
A histological section was annotated by a neuropathologist, delineating the tumour centre and the infiltration area, as well as, grey and white matter in the tumour-free sample.
In \cref{fig:SanityCheckAzimuthalCircSD} the \textit{csd($\varphi$)} in a fast and single-pass denoising with PDDN is compared against the LQ, HQ and SHQ instances as reference.
The denoising reduces the \textit{csd($\varphi$)} in both tumour-free and diseased samples similarly to HQ and SHQ data.
LQ instances show a high level of angular variability due to the intrinsic polarimetric acquisition noise and computational error propagation.
More homogeneous directional patterns are found in tumour-free white matter, with lower \textit{csd($\varphi$)} compared to grey matter, suggesting more organised fibre tracts, whilst crossing fibres increase the angular deviation.
The denoised lesion shows a clear difference between tumour centre and infiltration area, similarly to HQ and SHQ data.
A substantially higher \textit{csd($\varphi$)} is found for the tumour centre compared to the infiltration area, where higher variability suggests a higher degree of disruption of axonal fibres.
Conversely, the LQ instance showed limited separation of the tumour core from the background and the neighbouring structures.
The qualitative observations are reflected in the box-plot distributions, in line with the underlying tissue classes, even for higher polarimetric image quality standards.

\subsection{End-to-End Computational Performance}
\label{ExperimResults:ComputationalPerformance}

The end-to-end polarimetric processing pipeline accounts for single-shot denoising and parameters derivation.
Traditional multi-shot averaging techniques and derivation algorithms in \cite{Novikova2023Mueller} are considered as reference for HQ MPI. 
The performance is reported in \cref{tab:ComputationalPerformance} for a local patch and a full-scale image.
The comparison showed a substantial reduction in total processing time.
Our end-to-end pipeline achieved real-time performance ($<40$ms) for a tensor patch (size: $128\times 128\times16$) of polarimetric intensities.
This suggests that translation to \textit{in-vivo}, real-time MPI can already be achieved by focusing on a smaller field of view, with further optimisation necessary for full-frame images.
All computations were performed on a Linux Ubuntu 20.04 laptop, 16$\times$CPU at 2.6 GHz, 64 GB RAM, NVIDIA RTX A5000 GPU.

\begin{table}[!b]
    \centering
    \scriptsize
    \begin{tabular}{@{}c@{\hskip 0.5em}c@{\hskip 0.5em}c@{\hskip 0.5em}c|c@{\hskip 0.5em}c@{\hskip 0.5em}c@{}}
         & \multicolumn{3}{c}{Patch ($128 \times 128 \times 16$)} & \multicolumn{3}{c}{Full-scale Image ($512 \times 384 \times 16$)} \\
         \toprule \\[-2.5ex]
         & Denoising & Derivation & Total & Denoising & Derivation & Total \\
         \cite{Novikova2023Mueller} & 3.2$\pm$0.53s & 1.36$\pm$0.42s & 4.37$\pm$0.94s & 30.2$\pm$5.47s & 15.3$\pm$4.79s & 45.5$\pm$10.3s \\
         Ours & \textbf{15.4$\pm$0.17}ms & \textbf{23.4$\pm$4.5}ms & \textbf{38.8$\pm$4.67}ms & \textbf{0.55$\pm$0.08}s & \textbf{0.26$\pm$0.05}s & \textbf{0.81$\pm$0.06}s \\\\[-1em]
         \botrule
    \end{tabular}
    \normalsize
    \caption{End-to-end Computational Performance: pipeline processing time on same hardware. Time reported mean $\pm$ sd comparing our approach to reference. Best performance in bold. Reference: CPU-based implementation (Matlab R2021a). Ours: GPU (denoising) + CPUs (derivation) at full operational capacity. Multi-fold reduction of processing time, towards feasible real-time ($\leq 40$ ms) neurosurgical translations.}
    \label{tab:ComputationalPerformance}
\end{table}

\section{Discussion and Conclusions}
\label{DiscussionConclusions}

\begin{figure}[t]
    \centering
    \includegraphics[width=\textwidth]{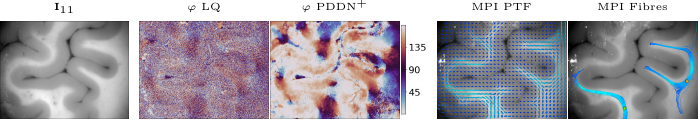}
    \caption{Denoised Polarimetric Tractography: tumour-free brain sample. LQ and denoised azimuth with recursive PDDN$^{+}$. MPI PTF: parameters mapped into an ellipsoidal model for tractography. Colours code for trace of ellipsoids eigenvalues. Fibres in seeding regions as in neurosurgical probing.}
    \label{fig:ProspectiveDenoisedPolarimTractography}
\end{figure}

In this feasibility study, we introduced a novel polarimetric denoising framework, with the goal of enabling high quality, high performance MPI for neurosurgery.
Developments combined our PDDN, for accurately enhancing images from short-time low-quality acquisitions, with a performance-optimised toolkit to efficiently derive parameters of clinical relevance.
The validation reported significantly improved image quality and achieved real-time performance for a local field of view.
The denosing accuracy was tested on multiple and diverse instances of human brain samples for different image restoration methods.
Our self-supervised PDDN yielded best rejection of the acquisition noise and limited the error propagation in the computational cascade, with comparable values to reference HQ data.
Whilst multi-shots averaging \cite{Novikova2023Mueller,Rodriguez-Nunez2022Polarimetric} produces reference MPI, it is incompatible with \textit{in-vivo} neurosurgery, where real-time feedback is needed.
Bypassing time-consuming MPI with enhanced image processing was first proposed in \cite{Li2020LearningBased,Yang2022DeepLearning}, where \mbox{U-Net}-like architectures (PCNN) denoised Mueller matrices derived from noisy, short-time acquisitions.
In \cite{Yang2022DeepLearning}, Mueller coefficients were denoised after training on large, paired, histological data, with inferential performance not yet compatible with real-time applications.
In our experiments PDDN over-performed PCNN and traditional denoising methods.
This is likely due to a combination of factors: the different nature of input polarimetric data, the type of noise, and the underlying probabilistic model.
While our model is specific to the considered human brain samples, and for the specific polarimetric acquisition conditions, differently from \cite{Li2020LearningBased,Yang2022DeepLearning}, the PDDN denoises source intensities corrupted by acquisition noise, with the Mueller coefficients being subsequently \textit{derived}.
The rationale behind adopting PDDN builds on empirical similarities between measured MPI acquisition noise, i.e. pseudo-Gaussian: symmetric, zero-mean, bell-shaped, with cumulative slightly deviating from the Normal reference, and the additive noise in the probabilistic formulation.
As denoising diffusion networks can generalise for complex distributions \cite{Ho2020Denoising}, we aimed to reduce MPI acquisition noise by generalising for contrast variability in biological structures with few filtering steps.
The initial calibration mitigated systematic errors in the polarisation states, however, different wavelengths and varying exposure time may introduce a non-linear intensity bias together with other specific human brain tissue structures (e.g. cortical regions in eloquent areas vs. deep-brain structures of corpus callosum, or other structures of the cerebellum), which may potentially alter the image contrast, structural patterns, values and noise distributions propagated in the Mueller derivations.
In this case, our specific model was able to generalise for the considered polarisation states, for the intensity bias, for fresh and formalin-fixed samples, and for tumour-free and neoplastic tissues, by preserving the underlying micro-structure after denoising.
Clear cortical white matter fibres orientations and comparable azimuth deviations were observed after denoising with respect to HQ and SHQ data for a representative tumour-free sample and a glioma lesion in \cref{fig:SanityCheckAzimuthalCircSD}.
Angular deviations were visible after denoising, in keeping with underlying tissue: the consistent drop in azimuth variability showed higher compression and reduced overlap among pathological regions, better than HQ and SHQ data.
Interestingly, PDDN was \textit{only} trained on HQ images, yet azimuth deviations were similar to SHQ data, suggesting high MPI quality is achievable with AI beyond conventional acquisition paradigms.
Prospectively, advanced configurations (PDDN$^{+}$) may enable neurosurgical fibre tracking with polarimetric tensor fields (PTF) in \cref{fig:ProspectiveDenoisedPolarimTractography}.
Future analyses will test multi-spectral denosing \textit{in-vivo}, accounting for motion and bleeding artifacts.
MPI instrumentation optimisations and image processing developments will be tailored on edge-computing solutions, for real-time wide-field MPI video streams.

\backmatter

\bmhead{Supplementary materials}
Representative polarimetric brain data employed in the validating dataset are available at: \url{https://osf.io/9ynmf/}.

\bmhead{Acknowledgments}
This work was supported by the Swiss National Science Foundation (SNSF) Sinergia Grant No. CRSII5$\_$205904 ``HORAO - Polarimetric visualization of healthy brain fiber tracts for tumor delineation during neurosurgery''.
Prototypes for high performance data capturing and visualisation were possible thanks to NVIDIA Clara AGX\textsuperscript{\texttrademark}, the Holoscan platform and guidance from the team.

\bibliography{PolarimetryBiblio.bib}

\end{document}